\documentclass[11pt]{article}
\usepackage{amssymb}
\usepackage{graphicx}
\usepackage{layout}
\usepackage{amssymb}
\newcommand{\fnref}[1]{~\ref{#1}}
\begin{document}


\title{The Sagnac Phase Shift suggested by the Aharonov-Bohm effect
for relativistic matter beams}
\author{Guido Rizzi$^{\S,\P}$  and Matteo Luca Ruggiero$^{\S,\P,}$\thanks{%
Dipartimento di Fisica del Politecnico di Torino, Corso Duca degli
Abruzzi 24, I-10129 Torino - Italy, tel. \ +390115647380, fax \
+390115647399}
\\
\small
$^\S$ Dipartimento di Fisica, Politecnico di Torino,\\
\small $^\P$ INFN, Sezione di Torino\\
\small E-mail guido.rizzi@polito.it, matteo.ruggiero@polito.it}
\maketitle

\begin{abstract}
The phase shift due to the Sagnac Effect, for relativistic matter
beams counter-propagating in a rotating interferometer, is deduced
on the bases of a a formal analogy with the the Aharonov-Bohm
effect. A procedure outlined by Sakurai, in which non relativistic
quantum mechanics and newtonian physics appear together with some
intrinsically relativistic elements, is generalized to a fully
relativistic context, using the Cattaneo's splitting technique.
This approach leads to an exact derivation, in a self-consistently
relativistic way, of the Sagnac effect. Sakurai's result is
recovered in the first order approximation.
\end{abstract}


\section{Introduction}\label{sec:intro}

\subsection{The early years}\label{ssec:early}
The story of the interferometrical detection of the effects of
rotation dates back to the end of the XIX century when, still in
the context of the ether theory, Sir Oliver Lodge\cite{lodge93}
proposed to use a large interferometer to detect the rotation of
the Earth. Subsequently\cite{lodge97} he proposed to use an
interferometer rotating on a turntable in order to reveal rotation
effects with respect to the laboratory frame. A detailed
description of these early works can be found in the paper by
Anderson {\it et al.}\cite{anderson94}, where the study of
rotating interferometers is analyzed in a historical perspective.
In 1913 Sagnac\cite{sagnac13} verified his early
predictions\cite{sagnac05}, using a rapidly rotating light-optical
interferometer. In fact, on the ground of classical physics, he
predicted the following fringe shift (with respect to the
interference pattern when the device is at rest), for
monochromatic light waves in vacuum, counter-propagating along a
closed path in a rotating  interferometer:
\begin{equation}
\Delta z=\frac{4{\bf \Omega \cdot S}}{\lambda c}  \label{eq:sagnac1}
\end{equation}
where  ${\bf \Omega }$ is the (constant) angular velocity vector of the turntable,  $%
{\bf S}$ is the vector associated to the area enclosed by the
light path, and $\lambda $ is the wavelength of light in vacuum.
The time difference associated to the fringe shift (\ref
{eq:sagnac1}) turns out to be
\begin{equation}
\Delta t=\frac{\lambda }{c}\Delta z=\frac{4{\bf \Omega \cdot
S}}{c^{2}} \label{eq:sagnac2}
\end{equation}

Even if his interpretation of these results was entirely in the
framework of the classical (non Lorentz!) ether theory, Sagnac was
the first scientist who reported an experimental observation of
the effect of rotation on spacetime, which, after him, was named
''Sagnac effect''. It is interesting to notice that the Sagnac
effect was interpreted as a disproval of the Special Theory of
Relativity (SRT) not only during the early years of relativity (in
particular by Sagnac himself), but, also, more recently, in the 90's by Selleri%
\cite{selleri96},\cite{selleri97}, Croca-Selleri\cite{croca99}, Goy-Selleri%
\cite{goy97}, Vigier\cite{vigier97}, Anastasovski {\it et
al.}\cite {anastasowski99}. However, this claim is incorrect: the
Sagnac effect can be explained completely in the framework of SRT,
 which allows  a  deeper insight into its very foundations. In fact,
it can be interpreted as an observable consequence of the
synchronization gap predicted by  SRT for non-time-orthogonal
physical frames (see Weber\cite {weber97}, Dieks\cite{dieks91},
Anandan\cite{anandan81}, Rizzi-Tartaglia\cite {rizzi98},
Bergia-Guidone \cite{bergia98},
Rodrigues-Sharif\cite{rodrigues01}). In particular, SRT predicts
the following proper time difference (as measured by a clock at
rest in the starting/ending point on the turntable) between light
beams counter-propagating in a ring interferometer
\begin{equation}
\Delta \tau=\frac{4\pi R^{2}\Omega }{c^{2}\left( 1-\frac{\Omega ^{2}R^{2}}{c^{2}%
}\right) ^{1/2}}  \label{eq:sagnac3}
\end{equation}
where $R$ is the radius of the ring.  Evidently, relation
(\ref{eq:sagnac3}) reduces to (\ref{eq:sagnac2}) in the first
order approximation (with respect to the small parameter
$\frac{\Omega R}{c}$).

Few years before Sagnac, Franz Harres\cite{harres12}, graduate
student in Jena, observed, for the first time but unknowingly, the
Sagnac effect during his experiments on the Fresnel-Fizeau drag of
light. However, only in 1914, Harzer\cite{harzer14} recognized
that the unexpected and inexplicable bias found by Harres was
nothing else than the manifestation of the Sagnac effect.
Moreover, Harres's observations also demonstrated that the Sagnac
fringe shift is unaffected by refraction: in other words, it is
always given by eq. (\ref{eq:sagnac1}), provided that $\lambda $
is interpreted as the light wavelength in a comoving refractive
medium. So, the Sagnac phase shift depends on the light
wavelength, and not on the velocity of light in the (comoving)
medium.

If Harres anticipated the Sagnac effect on the experimental ground, Michelson%
\cite{michelson04} anticipated the effect on the theoretical side.
Subsequently, in 1925, Michelson himself and
Gale\cite{michelson25} succeeded in measuring a phase shift,
analogous to the Sagnac's one, caused by the rotation of the
Earth, using a large optical interferometer.

The field of light-optical Sagnac interferometry had a revived
interest after the development of laser (see for instance the
beautiful review paper by Post\cite{post67}, where the previous
experiments are carefully described and their theoretical
implications analyzed). As a consequence, there was an increasing
precision in measurements and a growth  of technological
applications, such as inertial
navigation\cite{chow85}, where the "fiber-optical gyro"%
\cite{vali76} and the "ring laser"\cite{stedman97} are used.

\subsection{Universality of the Sagnac
Effect}\label{ssec:relativistic}

Until now, we have been speaking of the Sagnac effect for light
waves. However the effect has an universality which goes beyond
the nature of the interfering beams: this can be easily
demonstrated and understood in SRT.

The validity of eq. (\ref{eq:sagnac3}) for any couple of
counter-propagating electromagnetic beams is a very remarkable
feature of the Sagnac effect, and a first important indication of
its universality. In fact it shows that the effect depends  only
on the angular velocity of the turntable and on the path of the
beams on the turntable; on the contrary, it does not depend on the
light wavelength and on the presence of the (comoving) optical
medium.

However, the strongest claim from its universality comes from the
fact that the effect turns out to be exactly the same for any kind
of ''entities'' (such as electromagnetic and acoustic waves,
classical particles and electron Cooper pairs, neutron beams and
De Broglie waves and so on...) travelling in opposite directions
along a closed path in a rotating interferometer, with the same
(in absolute value) velocity with respect to the turntable. Of
course the ''entities'' take different times  for a complete
round-trip, depending on their velocity relative to the turntable;
\textit{but the difference between these times is always given by
eq. (\ref{eq:sagnac3})}. For matter entities, this time difference
can be obtained, for instance, using the relativistic law of
velocity composition (see Malykin\cite{malykin00} and
Rizzi-Ruggiero\cite{rizzi03}). So, the amount of the time
difference is always the same, both for matter and light waves,
independently of the physical nature of the interfering beams.

This astounding but experimentally well proved fact, is the most
important clue for preferring the special relativistic explanation
of the Sagnac effect. In fact, its ''universality'' cannot be
explained on the bases of the classical physics, but it can be
easily explained as a "geometrical effect" in spacetime, on the
bases of relativistic physics. In fact, in SRT, the crucial clue
leading to a geometrical (i.e. universal) explanation is the fact
that the time difference between any couple of ''entities''
exactly coincides with (twice) the synchronization gap predicted
 for non-time-orthogonal physical frames (the so-called
''time-lag'', see f.i. Anandan\cite{anandan81} and
Rizzi-Tartaglia\cite{rizzi98}).\bigskip

\subsection{Experimental tests and derivation of the Sagnac Effect}
\label{ssec:tests}

The Sagnac effect with matter waves has been verified
experimentally using Cooper pair\cite{zimmermann65} in 1965, using
neutrons\cite{atwood84} in 1984, using $^{40}Ca$ atoms
beams\cite{riehle91} in 1991 and using electrons, by
Hasselbach-Nicklaus\cite{hasselbach93}, in 1993. The effect of the
terrestrial rotation on neutron phase was demonstrated in 1979 by
Werner {\it et al.}\cite{werner79} in a series of famous
experiments.

The Sagnac phase shift has been derived, in the first order
approximation, in various ways by different authors (see the paper
by Hasselbach-Nicklaus quoted above, for discussion and further
references), often using an heterogeneous mixture of classical
kinematics and relativistic dynamics, or non relativistic quantum
mechanics and some relativistic elements.

An example of derivation of the Sagnac effect for material  beams,
which is based on this odd mixture of non-relativistic quantum
mechanics, newtonian mechanics and intrinsically relativistic
elements, was given in a well known paper by
Sakurai\cite{sakurai80}. Sakurai's derivation is based on a formal
analogy between the {\it classical }Coriolis force
\begin{equation}
{\bf F}_{Cor}=2m_{o}{\bf v}\times {\bf \Omega \;,}  \label{eq:coriolis1}
\end{equation}
acting on a particle of mass $m_{o}$  moving in a uniformly
rotating frame, and the Lorentz force
\begin{equation}
{\bf F}_{Lor}=\frac{e}{c}{\bf v}\times {\bf B}  \label{eq:lorentz1}
\end{equation}
acting on a particle of charge $e$ moving in a constant magnetic
field ${\bf B}$.

Let us consider a beam of charged particles split into two
different paths and then recombined. If $S$ is the surface domain
enclosed by the two paths, the resulting phase difference in the
interference region turns out to be:
\begin{equation}
\Delta \Phi =\frac{e}{c\hbar }\int_{S}{\bf B}\cdot {\rm d}{\bf S}
\label{eq:phasemag}
\end{equation}
Therefore, $\Delta \Phi $ is different from zero when a magnetic
field  exists \textit{inside} the domain enclosed by the two
paths, even if  the magnetic field felt by the
particles along their paths is zero. This is the well known Aharonov-Bohm\cite{aharonovbohm59} effect\footnote{%
In the case of the Aharonov-Bohm effect, the magnetic field
$\textbf{B}$ is zero along the trajectories of the particles,
while in the Sakurai's derivation, which we are going to
generalize, the angular velocity, which is the analogue of the
magnetic field for particles in a rotating frames, is not null:
therefore the analogy with the Aharonov-Bohm effect seems to be
questionable. However, the formal analogy can be easily recovered
when \textit{the flux} of the magnetic field, rather than the
magnetic field itself, is considered: this is just what we are
going to do (see Section \ref{sec:ab}, below).}.

By  formally substituting
\begin{equation}
\frac{e}{c}{\bf B}\rightarrow 2m_{o}{\bf \Omega }
\label{eq:subsak}
\end{equation}
Sakurai shows that the phase shift (\ref{eq:phasemag}) reduces to
\begin{equation}
\Delta \Phi =\frac{2m_{o}}{\hbar }\int {\bf \Omega }\cdot d{\bf S}
\label{eq:phaseomega}
\end{equation}
If ${\bf \Omega}$ is interpreted as the angular velocity vector of
the uniformly rotating turntable, and ${\bf S}$ as the vector
associated to the area enclosed by the closed path  along which
two counter-propagating material  beams travel, then eq.
(\ref{eq:phaseomega}) can be interpreted as the Sagnac phase shift
for the considered counter-propagating beams:
\begin{equation}
\Delta \Phi = \frac{2 m_o}{\hbar} {\bf \Omega} \cdot {\bf S}
\label{eq:phaseomega1}
\end{equation}
This result has been obtained using non relativistic quantum
mechanics. The time difference corresponding to the phase
difference (\ref{eq:phaseomega1}), turns out to be:
\begin{equation}
\Delta t=\frac{\Delta \Phi }{\omega }=\frac{\hbar }{E}\Delta \Phi =\frac{%
\hbar }{mc^{2}}\Delta \Phi = \frac{2m_o}{m c^2} {\bf \Omega} \cdot
{\bf S} \label{eq:deltat}
\end{equation}
Let us point out that eq. (\ref{eq:deltat}) contains,
un-consistently but unavoidably, some relativistic elements
($\hbar \omega =E=mc^{2}$). Of course in the first order
approximation, i.e. when the relativistic mass $m$ coincides with
the rest mass $m_{o}$ eq. (\ref{eq:deltat}) reduces to eq.
(\ref{eq:sagnac2}); that is, as we stressed before, a first order
approximation for the relativistic time difference
(\ref{eq:sagnac3}) associated to the Sagnac
effect\footnote{Formulas (\ref{eq:sagnac2}) and (\ref{eq:deltat})
differs by a factor 2: this depends on the fact that in eq.
(\ref{eq:sagnac2}) we considered the complete round-trip of the
beams, while in this section  we refer to a situation in which the
emission point and the interference point are diametrically
opposed.\label{fn:fact2}}.

\subsection{A generalization of the Sakurai's derivation}\label{ssec:general}

In this paper we are going to extend the simple ''derivation by
analogy'' used by Sakurai to a fully relativistic context. To this
end the Cattaneo's 1+3 splitting\cite{cattaneo},\cite{catt1},\cite
{catt2},\cite{catt3},\cite{catt4} will be adopted: it will enable
us to describe the geometrodynamics of the rotating frame in a
very transparent and powerful way. In particular, the Catteneo's
splitting allows to generalize the newtonian elements used by
Sakurai to a  relativistic context, in which also relativistic
quantum mechanics can be adopted. This new approach leads to a
derivation, in a self-consistent  way, of the relativistic Sagnac
time delay (\ref{eq:sagnac3}), whose first order approximation
coincides with Sakurai's result (\ref{eq:deltat}). Moreover,
contrary to Sakurai's claim (see footnote 7 of the paper quoted
above), in our derivation it is shown that the analogy between the
Sagnac phase shift and the Aharonov-Bohm phase shift holds also in
relativistic quantum mechanics.

\section{The  phase of quantum particles in electromagnetic field and
the Aharonov-Bohm Effect}\label{sec:ab}

Let us consider a quantum particle of (proper) mass $m_{o}$ and electric
charge $e$. If the particle is free, the associated Dirac equation is \cite
{sakurailibro}
\begin{equation}
\left( \gamma ^{\mu }\partial _{\mu }+\frac{m_{o}c}{\hbar }\right) \psi (x)=0
\label{eq:dirac1}
\end{equation}
where $\psi (x)$ is the spinorial wave function  which is the solution of (%
\ref{eq:dirac1}) and $x\equiv \left\{ x^{\mu }\right\} $ is a
point in spacetime\footnote{Let $(-1,1,1,1)$ be the signature of
spacetime; Greek indices run from 0 to 3, while Latin indices run
from 1 to 3.}.

In an electromagnetic field described by the 4-potential $A_{\mu }$ the
Dirac equation is obtained by the formal substitution $\partial _{\mu
}\rightarrow $ $\partial _{\mu }-i\frac{e}{\hbar c}A_{\mu }\,$, and the wave
equation becomes
\begin{equation}
\left[ \left( \gamma ^{\mu }\left( \partial _{\mu }-i\frac{e}{\hbar c}A_{\mu
}\right) +\frac{m_{o}c}{\hbar }\right) \right] \psi ^{\prime }(x)=0
\label{eq:dirac2}
\end{equation}
where $\psi ^{\prime }(x)$ is the spinorial solution of (\ref{eq:dirac2}).

According to  this formulation of the interaction between the
electromagnetic field and the particle, it can be shown that, if
$\psi (x)$ is a solution of a physical problem for the free
quantum particle according to (\ref{eq:dirac1}), the corresponding
solution for the interacting wave equation (\ref{eq:dirac2}) turns
out to be
\begin{equation}
\psi ^{\prime }(x)=\exp \left( i\frac{e}{\hbar c }\int^{x}A_{\mu
}(x^{\prime })dx^{\prime }{}^{\mu }\right) \psi (x)
\label{eq:psi1}
\end{equation}
One says that the $A_{\mu }$ field has produced a non-integrable
phase factor that depends on the past history of the particle,
which appears in (\ref{eq:psi1}) as the domain of
integration\footnote{This is a very general result, that applies
as well  to the Schr\"{o}dinger wave function of an interacting
non relativistic particle (see below).}.

\begin{figure}[top]
\begin{center}
\includegraphics[width=10cm,height=12cm]{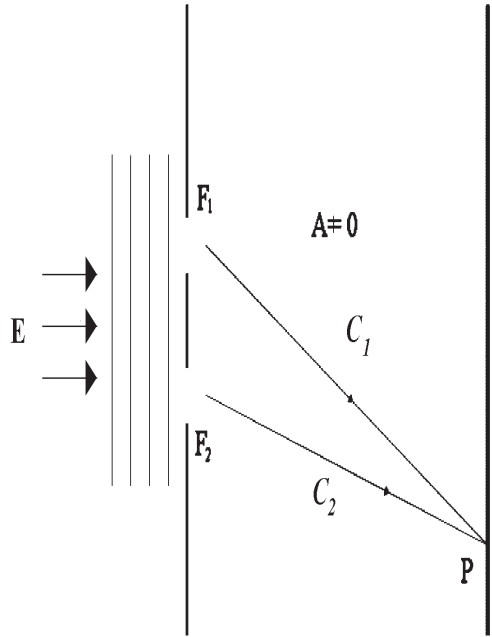}
\caption{\small A single coherent charged beam, originating in
$E$, is split into two parts (passing through the two slits $F_1$
and $F_2$) that propagate, respectively,  along the paths $C_1$
and $C_2$ (in the figure these paths are represented,
respectively, by  $EF_1P$ and $EF_2P$).  The beams travel in a
region where a vector potential $\mathbf{A}$ is present. In $P$,
the beams interfere and an additional phase shift is provoked by
the magnetic field.} \label{fig:ab1}
\end{center}
\end{figure}
\normalsize

This analysis leads to the existence of a remarkable phenomenon.\newline

Consider the two slits experiment (figure 1) and imagine that a
single coherent charged beam is split into two parts, which travel
in a region where only a magnetic field is present, described by
the 3-vector potential ${\bf A}$; then the beams are recombined to
observe the interference pattern. The phase  of the two wave
functions, at each point of the pattern, will be modified, with
respect to the case of free propagation (${\bf A}~=~0$), by
factors of the form given in (\ref{eq:psi1}), which depend on the
respective space trajectories. The magnetic potential-induced
phase shift has the form
\begin{equation}
\Delta \Phi=\frac{e}{c \hbar} \left(\int_{C_1} A_i
dx'^i-\int_{C_2} A_i
dx'^i \right)=\frac{e}{c\hbar }\oint_{C}{\bf A}\cdot {\rm d}{\bf r}=\frac{e}{%
c\hbar }\int_{S}{\bf B}\cdot {\rm d}{\bf S}  \label{eq:ab}
\end{equation}
where $C$ is the oriented closed curve, obtained as the sum of the
oriented paths $C_{1}$ and $C_{2}$ relative to each component of
the beam (in the physical space, see figure \ref{fig:ab1}). Eq.
(\ref{eq:ab}) expresses (by means of the Stoke's Theorem) the
phase difference in terms of the flux of the magnetic field across
the surface $S$ enclosed by the curve $C$.

\begin{figure}[top]
\begin{center}
\includegraphics[width=10cm,height=12cm]{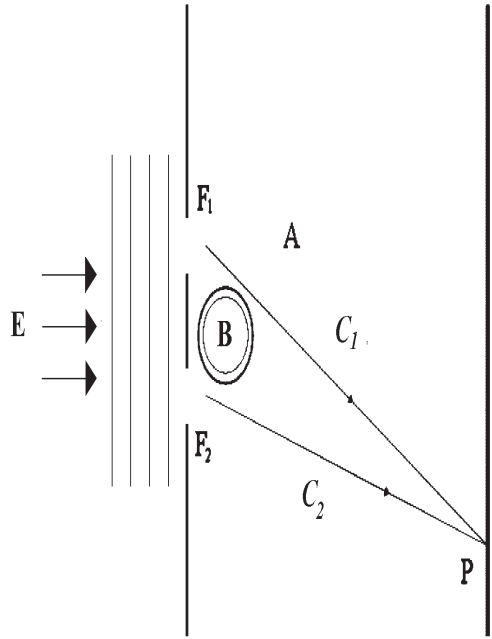}
\caption{\small A single coherent charged beam, originating in
$E$, is split into two parts (passing through the two slits $F_1$
and $F_2$) that propagate, respectively,  along the paths $C_1$
and $C_2$ (in the figure these paths are represented,
respectively, by  $EF_1P$ and $EF_2P$). Between the paths a
solenoid is present; the magnetic field $\mathbf{B}$ is entirely
contained inside the solenoid, while outside there is a constant
vector potential $\mathbf{A}$. In $P$, the beams interfere and an
additional phase shift, provoked by the magnetic field confined
inside the solenoid, is observed.} \label{fig:ab2}
\end{center}
\end{figure}

Aharonov and Bohm\cite{aharonovbohm59} applied this result to the
situation in which the two  split beams pass one on each side of a
solenoid inserted between the paths (see figure \ref{fig:ab2}).
Thus, even if the magnetic field ${\bf B}$ is totally contained
within the solenoid, and the beams pass through a ${\bf B}~=~0$
region, a resulting phase shift appears, since a non null magnetic
flux is associated to every closed path which encloses the
solenoid.\\

We need a relativistic wave equation in order to generalize the
Sakurai's "derivation by analogy" to a fully relativistic context.
However, Tourrenc\cite{tourrenc77} showed that no explicit wave
equation is demanded to describe the Aharonov-Bohm effect, since
its interpretation is a pure geometric one: in fact eq.
(\ref{eq:ab}) is independent of the very nature of the interfering
charged beams, which can be spinorial, vectorial or tensorial. In
particular, from a physical point of view,  spin has no influence
on the Aharonov-Bohm effect because there is no coupling with the
magnetic field which is confined inside the solenoid\footnote{If
the magnetic field is null, the Dirac equation is equivalent to
the Klein-Gordon equation, and this is the case of a situation
when a constant potential is present. Therefore,  in what follows
we shall just use eq. (\ref{eq:ab}) and we shall not refer
explicitly to any relativistic wave equation.}.

Things are different when a particle with spin, moving in a
rotating frame, is considered. In this case a coupling between the
spin and the angular velocity of the frame appears (this effect is
evaluated by Hehl-Ni\cite{hehl90} and Mashhoon\cite{mashhoon88}).
As a consequence, our formal analogy between  matter waves, moving
in a uniformly rotating frame and  charged beams, moving in a
region\footnote{In a non rotating frame.} where a constant
magnetic potential is present, holds only when the spin-rotation
coupling is neglected.

\normalsize

\section{Generalized Coriolis and Lorentz Forces}\label{sec:potentials}

In this section we shall introduce the generalized Coriolis and
Lorentz forces, which will permit us to extend to a pure
relativistic context the Sakurai's procedure which we outlined in
Sect. \ref{ssec:tests}

First of all, let us choose a physical frame, which is represented
in spacetime by a time-like congruence $\Gamma $ of world lines of
the particles constituting the 3-dimensional physical frame; let
\mbox{\boldmath $\gamma$}$(x)$ be the field of unit vectors
tangent to the world lines of the congruence $\Gamma $. Now, let
us choose a system of admissible coordinate so that the lines
$x^{0}=var$ coincide with the lines of $\Gamma $; according to
Cattaneo's terminology, such coordinates are said to be `adapted
to the physical frame' defined by the congruence $\Gamma $.

Being $g_{\mu \nu }\gamma ^{\mu }\gamma ^{\nu }=-1$, the
controvariant and covariant components of the $\gamma $-field are:
\begin{equation}
\left\{
\begin{array}{c}
\gamma ^{{o}}=\frac{1}{\sqrt{-g_{{oo}}}} \\
\gamma ^{i}=0
\end{array}
\right. \;\;\;\;\;\;\;\;\;\;\;\;\left\{
\begin{array}{c}
\gamma _{o}=\sqrt{-g_{{oo}}} \\
\gamma _{i}=g_{i{o}}\gamma ^{{o}}
\end{array}
\right.   \label{eq:gamma}
\end{equation}
The physical spacetime is a (pseudo) riemannian manifold
$\mathcal{M}$, and in each point $p\in {\cal M}$, the tangent
space $T_{p}$ can be split into the direct sum of two subspaces:
$\Theta _{p}$, spanned by $\gamma ^{\alpha } $, which we shall
call "local time direction" of the given frame, and $\Sigma _{p}$,
the 3-dimensional subspace which is supplementary (M-orthogonal)
with respect to $T_{p}$; $\Sigma _{p}$ is called "local space
platform" of the given frame. So, the tangent space can be written
as the direct sum
\begin{equation}
T_{p}=\Theta _{p}\oplus \Sigma _{p}  \label{eq:tangsum}
\end{equation}

A vector which belongs $T_{p}$ can be projected onto $\Theta _{p}$
and $\Sigma _{p}$ using, respectively, the \textit{time projector}
$\gamma _{\mu }\gamma _{\nu }$ and the \textit{space projector}
$\gamma _{\mu \nu }\doteq g_{\mu \nu }-\gamma _{\mu }\gamma _{\nu
} $,  which is interpreted as ''spatial metric tensor''. Then the
''transverse'' derivative operator $\tilde{\partial}_{\mu }\doteq
\partial _{\mu }-\gamma _{\mu }\gamma ^{{o}}\partial _{{o}}$ can
be introduced (even
if  we shall confine ourselves only to  stationary situations, in which $%
\partial _{o}\equiv 0$). Finally, let us introduce the space vortex
tensor of the congruence:
\begin{equation}
\tilde{\Omega}_{hk}\doteq \gamma _{o}\left[ \tilde{\partial}_{h}\left( \frac{%
\gamma _{k}}{\gamma _{o}}\right) -\tilde{\partial}_{k}\left( \frac{\gamma
_{h}}{\gamma _{o}}\right) \right]   \label{eq:vorticespaziale}
\end{equation}
and let  \mbox{\boldmath $\omega$}$(x)$ $ \in \Sigma _{p}$ be the
axial 3-vector associated to the space vortex tensor of the
congruence by means of the relation
\begin{equation}
\omega ^{i}\doteq \frac{c}{4}\varepsilon ^{ijk}\tilde{\Omega}_{jk}=\frac{c}{2%
}\varepsilon ^{ijk}\gamma _{o}\tilde{\partial}_{j}\left( \frac{\gamma _{k}}{%
\gamma _{o}}\right)   \label{eq:vettorevortice}
\end{equation}
where $\epsilon^{ijk} \doteq \frac{1}{\sqrt{det(\gamma
_{ij})}}\delta ^{ijk}$ is the Ricci-Levi Civita tensor, defined in
terms of the completely antisymmetric symbol $\delta ^{ijk}$ and
of the spatial metric tensor $\gamma_{ij}$.

The equation of motion of a particle, relative to this physical
frame, can be obtained by means of the Cattaneo's projection
technique. In Appendix \ref{sec:Aeqmoto}  the general form of this
equation is given, in coordinates adapted to the physical frame
(see   eqs. (\ref{eq:moto1}),
(\ref{eq:motogen}),(\ref{eq:motogen1}))

In particular, in eq. (\ref{eq:motogen1}), a term which depends on
the 'standard relative velocity' ${\bf v}${\bf \ }of the particle
appears. It can be thought of as a generalized Coriolis-like
force:
\begin{equation}
{\cal F}_{i}=2m({\bf v}\times \mbox{\boldmath $\omega$})_{i}
\label{eq:corgen}
\end{equation}
where  $m$ is the relativistic mass $m\doteq m_{o}\left( 1-\frac{v^{2}}{c^{2}%
}\right) ^{-\frac{1}{2}}$ of the particle.

Now, let us introduce the ''gravito-electric potential'' $\phi ^{G}$ and the
''gravito-magnetic potential'' $A_{i}^{G}$ defined by
\begin{equation}
\left\{
\begin{array}{c}
\phi ^{G}\doteq -c^{2}\gamma ^{o} \\
A_{i}^{G}\doteq c^{2}\frac{\gamma _{i}}{\gamma _{o}}
\end{array}
\right.   \label{eq:gengravpot}
\end{equation}

In terms of these potentials, the vortex 3-vector ${\bf \omega
}^i$ is expressed in the form

\begin{equation}
\omega ^{i}=\frac{1}{2c}\varepsilon ^{ijk}\gamma _{o}\left( \tilde{\partial}%
_{j}A_{k}^{G}\right)   \label{eq:omega1}
\end{equation}
Alternatively, it can be written in the form
\begin{equation}
\omega ^{i}=\frac{1}{2c}\gamma _{o}\left( \widetilde{{\bf \nabla
}}\times {\bf A}_{G}\right) ^{i}\doteq \frac{1}{2c}\gamma
_{o}B_{G}^{i} \label{eq:omega2}
\end{equation}
where we implicitly defined the ''gravito-magnetic'' field
\begin{equation}
B_{G}^{i}\doteq \left( \widetilde{{\bf \nabla }}\times {\bf A}_{G}\right)
^{i}  \label{eq:gengravmag}
\end{equation}

In terms of this field, the velocity-dependent force (\ref{eq:corgen})
becomes
\begin{equation}
{\cal F}_{i}=m\gamma _{o}\left( \frac{{\bf v}}{c}\times {\bf B}_{G}\right)
_{i}  \label{eq:genlorentz}
\end{equation}
which has the form of a ''gravito-magnetic'' Lorentz force.

Notice that the Coriolis-like force (\ref{eq:corgen}) transforms into the
Lorentz-like force (\ref{eq:genlorentz}) with the formal substitution
\begin{equation}
2m\mbox{\boldmath $\omega$}\rightarrow \frac{m\gamma _{o}}{c}{\bf
B}_{G}  \label{eq:sub}
\end{equation}

\section{Sagnac effect for matter waves}\label{sec:sagnac}

Now we want to apply the formal analogy described in the previous
section to the phase shift induced by rotation on a beam of
massive particles which, after being split, propagate in two
opposite directions along the rim of a rotating disk. When they
are recombined, the resulting phase shift is the manifestation of
the Sagnac effect.

To this end, let us consider the analogue of the phase shift
(\ref{eq:ab}) for the gravito-magnetic field introduced before
\begin{equation}
\Delta \Phi =\frac{2m\gamma _{o}}{c\hbar }\oint_{C}{\bf A}^{G}\cdot {\rm d}%
{\bf r}=\frac{2m\gamma _{o}}{c\hbar }\int_{S}{\bf B}^{G}\cdot {\rm d}{\bf S}
\label{eq:gab}
\end{equation}
which is obtained on the bases of the formal analogy between eq.
(\ref{eq:genlorentz}) and  the  magnetic force
(\ref{eq:lorentz1}):
\begin{equation}
\frac{e}{c}\mathbf{B} \rightarrow \frac{m \gamma_o}{c}
\mathbf{B}^G \label{eq:sub2}
\end{equation}

To evaluate the phase shift (\ref{eq:gab}) we must consider the
congruence which describes the rotating frame in spacetime. In
particular, the space vectors belong to the (tangent bundle to
the)  ''relative space'' of the disk, which is the only space
having an actual physical meaning from an operational point of
view, and it is identified as the physical space of the rotating
platform\cite{rizzi02}.

Hence, in  the chart $(x^{0},x^{1},x^{2},x^{3})=(ct,r,\vartheta
,z)$ adapted to the rotating frame, the covariant components of
the metric tensor turn out to be\cite{rizzi02}:
\begin{equation}
g_{\mu \nu }=\left(
\begin{array}{cccc}
-1+\frac{\Omega ^{2}{r}^{2}}{c^{2}} & 0 & \frac{\Omega {r}^{2}}{c} & 0 \\
0 & 1 & 0 & 0 \\
\frac{\Omega {r}^{2}}{c} & 0 & {r}^{2} & 0 \\
0 & 0 & 0 & 1
\end{array}
\right)  \label{eq:metric}
\end{equation}
where $\Omega $ is the (constant) angular velocity of rotation of
the disk with respect to the laboratory frame. As a consequence,
the non null components of the vector field \mbox{\boldmath
$\gamma$}$(x)$, evaluated on the trajectory $R=const$ along which
both beams propagate, are:
\begin{equation}
\left\{
\begin{array}{c}
\gamma ^{o} \doteq \frac{1}{\sqrt{-g_{{oo}}}} = \gamma  \\
\gamma _{o} \doteq \sqrt{-g_{{oo}}} = \gamma ^{-1} \\
\gamma _{\vartheta}\doteq g_{\vartheta o} \gamma
^{{o}}=\frac{\gamma \Omega R^{2}}{c}
\end{array}
\right. \label{eq:gammas}
\end{equation}
where $\gamma =\left( 1-\frac{\Omega ^{2}R^{2}}{c^{2}}\right) ^{-\frac{1}{2}%
} $.

So, for the gravitomagnetic potential we obtain
\begin{equation}
A_{\vartheta }^{G} \doteq c^{2}\frac{\gamma _{\vartheta}}{\gamma
_{o}} =\gamma ^{2}\Omega R^{2}c \label{eq:vectpot}
\end{equation}

As a consequence, the phase shift (\ref{eq:gab}) becomes
\begin{equation}
\Delta \Phi =\frac{2m}{c\hbar \gamma }\int_{0}^{2\pi }A_{\vartheta
}^{G}d\vartheta =\frac{2m}{c\hbar \gamma }\int_{0}^{2\pi }\left( \gamma
^{2}\Omega R^{2}c\right) d\vartheta =4\pi \frac{m}{\hbar }\Omega R^{2}\gamma
\label{eq:deltaphigab}
\end{equation}

According to Cattaneo's terminology, the proper time is the
"standard relative time"  for an observer
on the rotating platform; so the proper time difference corresponding to (%
\ref{eq:deltaphigab}) is obtained according to

\begin{equation}
\Delta \tau = \frac{\Delta \Phi}{\omega}= \frac{\hbar}{E} \Delta
\Phi = \frac{\hbar}{mc^2}\Delta \Phi \label{eq:dtaudphi}
\end{equation}
and it turns out to be
\begin{equation}
\Delta \tau =4\pi \frac{\Omega R^{2}\gamma }{c^{2}} \equiv
\frac{4\pi R^2 \Omega}{c^2 \left( 1-\frac{\Omega ^{2}R^{2}}{c^{2}}
\right)^{1/2}} \label{eq:deltatau}
\end{equation}
which agrees with the proper time difference (\ref{eq:sagnac3})
due to the Sagnac effect, which, as we pointed out in subsection
\ref{ssec:relativistic}, corresponds to the time difference for
any kind of matter entities counter-propagating in a uniformly
rotating disk. As we stressed before,  this time difference does
not depend on the  standard relative velocity of the particles and it is exactly twice the {\it time lag }%
due to the synchronization gap arising in a rotating frame.

The phase shift can be expressed also as a function of the area
$S$ of the surface enclosed by the trajectories:
\begin{equation}
\Delta \Phi = 2 \beta^2 S \Omega \frac{m}{\hbar} \frac{\gamma^2}{\gamma-1}= 2%
\frac{m}{\hbar} S \Omega \left (\gamma+1 \right)
\label{eq:phiarea}
\end{equation}
where $\beta \doteq \frac{\Omega R}{c}$ and
\begin{equation}
S=\int_0^R \int_0^{2\pi} \frac{r dr d\vartheta}{\sqrt{1-\frac{\Omega^2 r^2}{%
c^2}}}=2\pi \frac{c^2}{\Omega^2}\left( 1-\sqrt{1-\frac{\Omega^2 R^2}{c^2}}%
\right) =2\pi \frac{c^2}{\Omega^2}\left(\frac{\gamma-1}{\gamma} \right)
\label{eq:area1}
\end{equation}
We notice that (\ref{eq:phiarea}) reduces to
(\ref{eq:phaseomega1})\footnote{Apart a factor 2, whose origin has
been explained in the footnote \fnref{fn:fact2} in Section
\ref{ssec:tests}.}, only in first order approximation with respect
to $\frac{\Omega R}{c}$, i.e. when $\gamma \rightarrow 1$: the
formal difference between (\ref{eq:phiarea}) and
(\ref{eq:phaseomega1}) is due to the non Euclidean features of the
relative space (see Rizzi-Ruggiero\cite{rizzi02} for further
details).

\section{Conclusions}\label{sec:conclusion}

The Sagnac phase shift for matter waves in a uniformly rotating
interferometer has been deduced, by means of a formal analogy with
the magnetic potential-induced phase shift for charged particles
travelling in a region where a constant vector potential is
present.

The formal analogy outlined by Sakurai, which explains the effect
of rotation using a "ill-assorted" mixture of non-relativistic
quantum mechanics, newtonian mechanics (which are
Galilei-covariant) and intrinsically relativistic
elements\footnote{Indeed, the lack of self-consistency, due to the
use of this odd mixture, is present not only in Sakurai's
derivation, but also in all the known approaches based on the
formal analogy with the Aharonov-Bohm effect.} (which are
Lorentz-covariant), has been extended to a fully relativistic
treatment, using the 1+3 Cattaneo's splitting technique. The space
in which waves propagate has been recognized as the relative space
of a rotating frame.

Using  this splitting technique,  we have generalized the
newtonian elements used by Sakurai to a fully  relativistic
context where we have been able to adopt relativistic quantum
mechanics. In this way, we have obtained   a derivation of the
relativistic Sagnac time delay (whose first order approximation
coincides with Sakurai's result)  in a self-consistent way.

\appendix

\section{Appendix: Equation of motion in an arbitrary physical  frame}\label{sec:Aeqmoto}

Given an arbitrary physical frame, the space projection (i.e. its
"standard relative formulation") of the equation of motion
\begin{equation}
\frac{Dp^{\alpha }}{d\tau }=F^{\alpha } \label{eq:eq:moto4d}
\end{equation}
of a particle in the external field described by the 4-vector
$F^{\alpha }$ turns out to be\footnote{The field $F^{\alpha }$
includes the possible constraints.}:
\begin{equation}
\frac{\hat{D}pi}{dT}=m(G_{i}^{\prime }+G_{i}^{\prime \prime })+F_{i}
\label{eq:moto1}
\end{equation}
where $T$ is the "standard relative time"; $\frac{\hat{D} \ }{d
T}$ is a suitable derivative operator; $p_{i},G_{i}^{\prime
},G_{i}^{\prime \prime }$ $F_{i}$ are
"relative" space vectors and $m$ is the relativistic mass $m~\doteq~m_{o}~\left( 1-\frac{v^{2}}{c^{2}%
}\right) ^{-\frac{1}{2}}$ of the particle, in terms of its
"standard relative velocity" $\mathbf{v}$(see Cattaneo
\cite{cattaneo},\cite{catt1},\cite
{catt2},\cite{catt3},\cite{catt4}. In particular in terms of the
potentials
\begin{equation}
\left\{
\begin{array}{c}
\phi _{G}\doteq -c^{2}\gamma ^{o} \\
A_{_{G}i}\doteq c^{2}\frac{\gamma _{i}}{\gamma _{o}}
\end{array}
\right. \label{eq:potentialgem}
\end{equation}
we can write
\begin{equation}
G_{i}^{\prime }=-\left( -\widetilde{\partial} _{i}\phi
_{G}-\partial _{o}A_{Gi}\right)
\end{equation}
which can be interpreted as a gravito-electric field:
\begin{equation}
E_{Gi}\doteq -\left( -\widetilde{\partial} _{i}\phi _{G}-\partial
_{o}A_{Gi}\right)  \label{eq:egen}
\end{equation}
\newline
Moreover, considering that
\begin{equation}
G_{i}^{\prime \prime } = 2\varepsilon _{ijk}\omega ^{k}v^{j}=\sqrt{%
det(\gamma _{ij})}\delta _{ijk}\omega ^{k}v^{j}
\end{equation}
and the definition of the gravitomagnetic field (\ref{eq:omega2}), the
equation of motion (\ref{eq:moto1}) can be written in the form
\begin{equation}
\frac{\hat{D}p_{i}}{dT}=mE_{Gi}+m\left( \frac{{\bf v}}{c}\times {\bf B}%
_{G}\right) _{i} + F_i  \label{eq:motogen}
\end{equation}
which is similar to the equation of motion of a particle acted
upon by a Lorentz force and an external field. Alternatively, we
can rewrite (\ref{eq:motogen})  using the rotation vector
\mbox{\boldmath $\omega$}
\begin{equation}
\frac{\hat{D}p_{i}}{dT}=mE_{Gi}+2m\left( {\bf v}\times
\mbox{\boldmath $\omega$}\right) _{i} + F_i  \label{eq:motogen1}
\end{equation}
where the Coriolis-like force ${\cal F}_{i}=2m({\bf v} \times
\mbox{\boldmath $\omega$})_{i}$  has been evidenced.

Although the most popular time+space splitting is the 1+3 ADM
splitting of Arnowitt, Deser, Misner\cite{ADM} (see
also\cite{MTW}), the simplicity of eqs.
(\ref{eq:motogen}),(\ref{eq:motogen1}) and their formal analogy
with the "classical" equation of motion, make the Cattaneo
splitting more suitable for our purposes.

For a  modern formulation of the Cattaneo's splitting, and its
relations with ADM splitting, see, for instance Jantzen\textit{et
al. }\cite{jantzen01a}, and the references therein.

\end{document}